\documentclass{emulateapj}
\usepackage[dvips]{color}

\usepackage{graphicx}
\slugcomment{Draft, \today}

\def\be{\begin{equation}}
\def\ee{\end{equation}}

\def\BH{_{\rm BH}}
\def\bh{_{\rm BH}}

\def\df{_{\rm df}}

\def\h{_{\rm h}}
\def\H{_{\rm H}}

\def\p{_{\rm p}}

\def\kms{{\rm\,km\,s^{-1}}}
\def\pc{{\rm\,pc}}
\def\kpc{{\rm\,kpc}}

\def\yr{{\rm\,yr}}
\def\Myr{{\rm\,Myr}}
\def\msun{{\rm\,M_\odot}}

\shorttitle{Kinematic distribution of the young stars in the GC}
\shortauthors{Yu, Lu, \& Lin}

\begin{document}
\title{On the origin of kinematic distribution of the sub-parsec 
young stars in the Galactic center}
\author{Qingjuan Yu$^{1,}$\footnotemark[3], Youjun Lu$^1$, \& 
D.~N.~C. Lin$^{1,2}$}
\affil{$^1$Department of Astronomy and Astrophysics, University of
           California, Santa Cruz, CA 95064, USA \\
       $^2$Kavli Institute of Astronomy \& Astrophysics, Peking
           University, Beijing, China}
\footnotetext[3]{Also a Hubble Fellow at the Department of Astronomy,
                 University of California at Berkeley, Berkeley, CA 94720.}
\email{yqj, lyj, lin@ucolick.org}

\begin{abstract}

Observations indicate the presence of a massive black hole in the Galactic
center. Within a half-parsec from the Galactic center, there is a population of
coeval young stars which appear to reside in a coherent disk. Surrounding this
dynamically-cool stellar system, there is a population of stars with a similar
age and much larger eccentricities and inclinations relative to the disk.  We
propose a hypothesis for the origin of this dynamical dichotomy.  Without
specifying any specific mechanism, we consider the possibility that both
stellar populations were formed within a disk some 6$\pm$2 Myr ago. But this
orderly structure was dynamically perturbed outside-in by an intruding
object with a mass $\sim 10^4 M_\odot$, which may be an intermediate-mass
black hole or a dark stellar cluster hosting an intermediate-mass black hole.
We suggest that the perturber migrated inward to $\sim0.15-0.3\pc$
from the Galactic center as a consequence of orbital decay under the action of
dynamical friction. Along the way, it captured many stars in the outer disk
region into its mean-motion resonance, forced them to migrate with it, closely
encountered with them, and induced the growth of their eccentricity and
inclination.  But stars in the inner regions of the disk retain their initial
coplanar structure. Quantitatively, a perturber on a low-inclination or
overhead orbit to the disk plane can reproduce the observed kinematic structure
of these young stars.  But this process is unlikely to produce the
controversial two-disk structure.  We predict that some of the inclined and
eccentric stars surrounding the disk may have similar Galactocentric semimajor
axis. Future precision determination of their kinematic distribution of these
stars will not only provide a test for this hypothesis but also evidences for
the presence of an intermediate-mass black hole or a dark cluster at the
immediate proximity of the massive black hole at the Galactic center.
\end{abstract}


\maketitle

\section{Introduction}\label{sec:intro}

Recent observations show that a large number of early-type stars,
either O/W-R stars or less massive B stars (totally about 90), reside
within the central half-parsec region of the Galactic center (GC)
\citep[and references therein]{Genzel03,Ghez03,Ghez05,Paumard06}.
These young stars have some distinctive features \citep{LB03,Genzel03,
Paumard06}: (a) About 26 stars
are found to reside in a well-defined and moderately thin disk, which
rotates clockwise in projection and has a scale height-to-radius ratio
$h/r\sim 0.12\pm 0.03$. According to \citet{Paumard06}, about a dozen 
stars reside in a less well-defined counterclockwise rotating disk and
these two disks are orientated at a large angle ($\sim 110\arcdeg$)
with each other. Note that the existence of the counterclockwise
rotating disk is still in controversy; nevertheless, observations
indicate that the motion of some stars (probably located at the outer
region, see item c below) is non-coplanar with respect to that of
stars in the clockwise rotating disk. (b) The stars in the clockwise
rotating disk are on low-eccentricity orbits (with eccentricity $e\sim
0.2-0.4$ on average). In contrast, many stars in the counterclockwise
disk appear to be on high-eccentricity orbits (say, $e\sim 0.8$). (c)
The clockwise rotating disk is compact and has a sharp inner edge at
$\sim 1\arcsec$ ($\sim 0.04\pc$ with an assumed GC's heliocentric
distance of $8\kpc$), while the counterclockwise component has a
ring-like structure which is located further out at $\sim 4\arcsec$
($\sim 0.15\pc$). (d) The stars in the ``two disks'' are coeval with
an age of $6\pm 2\Myr$ and
probably formed within a time span of $\la 2\Myr$.
(e) A cluster of main-sequence B stars ($4\msun\leq
M_*\leq 15\msun$) or the so-called ``S-cluster'' exists in the inner
$\sim 0.04\pc$ region, which has a spatially isotropic distribution
and is distinct from the more massive sub-parsec ($\sim 0.04-0.5\pc$)
young stars.  Some of these S-stars have high eccentricities (with $e$
up to 0.9--1) while others have moderate eccentricities.  The S-star
orbits have provided strong evidence for the existence of a massive
black hole (MBH; with mass $M\bh\simeq 3.6\times 10^6 \msun$) in the
GC \citep[and references therein]{Schodel02,Ghez05, Eisenhauer05}. 
The young stars in the GC, together with the central MBH, provide an
interesting dynamical system to study.  Specifically, the dynamical
architecture of the GC resembles that of the solar
system, which is composed of a thin disk of major planets and a thick
population of minor planets orbiting around the Sun.  As studies of
the dynamics in our solar system have provided us considerable
insights on the formation and evolution of the Sun and its surrounding
planets, investigations on the dynamical system in the GC may also
provide us insights into the structure, and the formation and
evolution of the central MBH and stars in the nucleus of our own Galaxy
and further in general galactic nuclei.

The coevality of the sub-parsec young stars suggests that they have a
common origin. Since the tidal force of the central MBH may prevent
formation of young stars through the collapse of self-gravitating cold
molecular gas clouds in the vicinity of the MBH \citep[e.g.,][]
{Sanders92, Morris93}, other hypothesis have been proposed to explain
the formation and youth of the stars, including some non-conventional
{\it in situ} formation scenario and in-spiraling young star cluster
scenario.  The {\it in situ} formation scenario is based on the
assumption that the young stars were formed via the onset of
gravitational instability and fragmentation in a hypothetical
accretion disk around the MBH which no longer exist today. With a
sufficiently large gas surface density, the self-gravity of the
perturbation may overcome the impediment of the MBH's tidal force
\citep[e.g.,][]{LB03,Genzel03,Goodman03,Levin07}. In the in-spiraling star
cluster scenario, it is assumed that the young stars were originally
formed in a dense star cluster outside the central half-parsec and
transported to their present location by the effect of dynamical
friction \citep{Gerhard01}. An intermediate-mass black hole (IMBH; a
few thousand solar mass) may be required to stabilize the cluster core
against its tidal disruption by the MBH before they move to the
central half-parsec region \citep{HM03,Kim04,GR05,MPZ03}. Note that the
existence of stellar disks in observations itself is not sufficient to
differentiate these competing scenarios since a young stellar disk may
form in the vicinity of the MBH in both of the scenarios.  (For a
review and more discussions on the pros and cons of these scenarios,
see \citealt{Genzel03,Ghez05,Paumard06,Alexander05}.)

Besides their young age, the peculiar orbital distribution of these stars also
provides clues to their formation mechanism and the dynamical structure in the
GC.  It can also be used to differentiate various models.  For example, in the
in-spiraling star cluster scenario, the interactions between the central IMBH
and the stripped young stars have been studied by \citet{Levin05} and
\citet{BH06}.  Although they were able to simulate the orbital features of
stars on the thin disk, they had difficulties to account for the kinematic
properties of those stars with eccentric orbits in the counterclockwise
rotating disk.
In this paper, we propose a dynamical model to account for the orbital
distribution of the sub-parsec young stars in the GC. Rather than
simultaneously investigating the formation mechanism of these young stars and
their orbital distribution, we assume that all the sub-parsec young stars were
formed in a clockwise rotating disk initially (e.g., due to instability or
fragmentation of a massive accretion disk) with small eccentricities.  We show
in \S~\ref{sec:timescales} that, an isolated stellar disk by itself is unlikely
to evolve from the presumed circular orbits in a thin disk to the currently
observed multi-component orbital configuration.  This inference is based on the
determination that the timescale (e.g., the two-body relaxation timescale and
the resonant relaxation timescale) for a single stellar disk to relax with
respect to the background stars is usually substantially longer than the age of
the young stars.  Thus, the life-long eccentricity growth of the disk stars due
to interactions among the stars is insignificant (not substantially larger than
0.2; \citealt{ABA07}).  Therefore, the existence of young stars with high
eccentricities and high inclination angles relative to the inner stellar disk
requires an explanation, at least in the disk formation scenario of the young
stars in the GC.  

It is entirely plausible that, under the action of dynamical friction,
a (dark) star cluster and/or an IMBH occasionally sink into the
proximity of the GC.  The (dark) star cluster (which may also have a
central IMBH) or the IMBH plays the role of a massive perturber (or
intruder), spiraling inward, inducing asymmetric perturbation of the
gravitational potential and exerting torques on the disk stars.
Interactions between the perturber and the disk stars, through either
close encounters or some resonant interactions, may change the orbital
distribution of the young stars. (In contrast to the in-spiraling
young cluster hypothesis, the primary role of the cluster
is to dynamically excite the eccentricity and inclination of the stars
along its path rather than to directly deliver its own young stars to
the GC.)  In this paper, we develop dynamical models in order to
examine whether the single stellar disk initially formed may evolve
into the current observed orbital configuration, especially the
eccentricity and inclination distribution, during and after the
passage of an inward migrating massive perturber.

This paper is organized as follows. In \S~\ref{sec:cusp}, we review
the components observed in the GC. In \S~\ref{sec:timescales}, we list
some relevant dynamical timescales in the GC.  In \S~\ref{sec:model},
we describe dynamical models of gravitational interaction between the
young stars in a primary disk and a massive perturber migrating inward
from outside the stellar disk in the GC. We present the results of a
set of numerical simulations of the dynamical models. Then we compare
the model results with observations. Finally, conclusions are given in
\S~\ref{sec:conclusion}.

\section{Components in the GC} \label{sec:cusp}

We consider three components which contribute to the gravitational potential
near the center of the Galaxy: a central MBH, a stellar cusp of old stars, and
a population of young stars mentioned in \S~\ref{sec:intro}.

The MBH has a mass $M\bh\simeq 3.6\times 10^6\msun$
\citep{Ghez05,Eisenhauer05}.  The radius of the central MBH's sphere of
influence is defined by
 \be a\H\equiv \frac{GM\BH}{\sigma^2}
        \simeq 1.6\pc~m\bh
\left(\frac{100\kms}{\sigma}\right)^2,
\label{eq:aH}
\ee
where $G$ is the gravitational constant, $\sigma$ is the one-dimensional
velocity dispersion, and $m\bh=M\bh/(3.6\times 10^6\msun)$ in the GC.

The stellar cusp surrounding the young stars is mainly composed of old stars
and stellar remnants including stellar-mass black holes and neutron stars.
The mass density of the stellar cusp in the GC can be described by a power law
\citep{Genzel03}, i.e., 
\be
\rho(r)=\rho_0\left(\frac{r}{r_0}\right)^{-\alpha_i},\ \ \ i=1, 2, 
\ee 
where $\rho_0=1.2\times 10^6 \msun\pc^{-3}$, $r_0=0.4\pc$,
$\alpha_1=1.4\pm 0.1$ for $r<r_0$ and $\alpha_2=2.0\pm 0.1$ for $r>r_0$. 
This mass density gives the enclosed stellar mass in the inner cusp by
\begin{eqnarray}
M_*(<r)& = &\frac{4\pi\rho_0 r_0^3}{3-\alpha_1}
            \left(\frac{r}{r_0}\right)^{3-\alpha_1} \nonumber \\
       & = &\epsilon M\bh f_1(x),  \ \ \ if\ \ r<r_0 \label{eq:masscusp} 
\end{eqnarray}
and 
\begin{eqnarray}
M_*(<r)& = &\frac{4\pi\rho_0 r_0^3}{3-\alpha_1}+ \frac{4\pi\rho_0 r_0^3}{3-
            \alpha_2}\left[\left(\frac{r}{r_0}\right)^{3-\alpha_2}-1\right]
            \nonumber \\
       & = &\epsilon M\bh f_2(x),  \ \ \ if\ \ r>r_0, 
\end{eqnarray}
where $\epsilon=\frac{4\pi\rho_0 r_0^3}{3-\alpha_1}/M\bh\sim
0.16$, $x=r/r_0$, $f_1(x)= x^{3-\alpha_1}$ and $f_2(x)=
\frac{3-\alpha_1}{3-\alpha_2}x^{3-\alpha_2}+
\frac{\alpha_1-\alpha_2}{3-\alpha_2}. $ The gravitational potential due 
to the cusp stars is
\be 
\Phi_*(r)=-\frac{GM_*(<r)}{r}+\frac{(3-\alpha_1)\epsilon
           GM\bh}{r_0}\ln\left(\frac{r} {r_0}\right), 
\label{eq:potencuspin} 
\ee for $r>r_0$ and 
\be
\Phi_*(r)=-\frac{GM_*(<r)}{r}-\frac{(3-\alpha_1)\epsilon
           GM\bh}{(2-\alpha_1)r_0}
           \left[1-\left(\frac{r}{r_0}\right)^{2-\alpha_1}\right],
\label{eq:potencuspout} 
\ee 
for $r<r_0$.  The total potential due to the MBH and the cusp stars is
$\Phi(r)=-\frac{GM\bh}{r}+\Phi_*(r)$.  Since the young stars interested in this
paper mainly reside within the central half parsec and $r_0=0.4\pc$, for
simplicity, we adopt $\alpha_2=\alpha_1\sim 1.4$ in our calculations below and
use equations (\ref{eq:masscusp}) and (\ref{eq:potencuspin}) even for $r>r_0$.
Our main results will not be affected by slightly changing the value of the
cusp slope.

\section{Some relevant dynamical timescales in the GC}\label{sec:timescales}

In this section, we estimate some dynamical timescales in the GC,
which may be used for references or to justify the approximations in
the dynamical models described in \S~\ref{sec:model}.

\begin{figure} \epsscale{1.0} 
\plotone{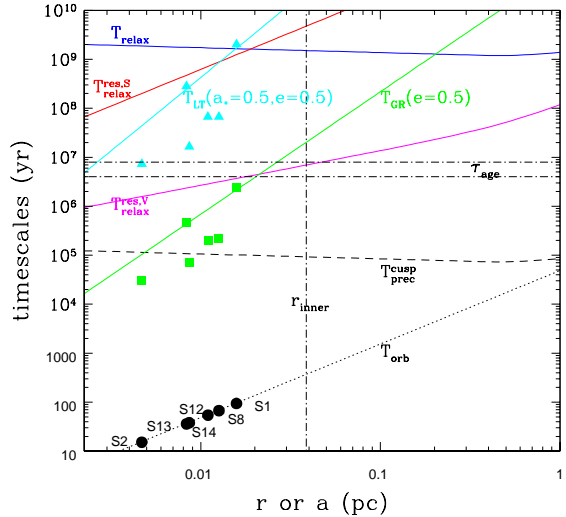}
\caption{Relevant timescales for the young stars in the GC as a function
of Galactocentric raidus $r$ or semimajor axis of a stellar orbit $a$: the
two-body relaxation timescale $T_{\rm relax}$ (blue solid line), the
scalar-resonant relaxation timescale $T^{\rm res,S}_{\rm relax}$ (red
solid line), the vector-resonant relaxation timescale $T^{\rm
res,V}_{\rm relax}$ (magenta solid line), the Lense-Thirring
precession timescale $T_{\rm LT}$ by assuming the dimensionless spin
parameter of the MBH $a_*=0.5$ and the eccentricity of the stars
$e=0.5$ (cyan solid line), the apsidal precession timescale due to the
general relativity correction $T_{\rm GR}$ with assuming $e=0.5$
(green solid line); the apsidal precession timescale due to the
stellar cusp $T^{\rm cusp}_{\rm prec}$ (dashed line); and the orbital
periods of the young stars $T_{\rm orb}$ (dotted line). For
comparison, the age of those young stars $\tau_{\rm age}\sim 6\pm
2\Myr$ and the inner radius of the clockwise disk $r_{\rm inner}\sim
1\arcsec\sim 0.04\pc$ are marked by the dot-dashed lines.  For
reference, the relevant timescales for several S-stars
\citep{Eisenhauer05,Ghez05} with measured orbital parameters are also
marked in this Figure (filled black circles: orbital periods; cyan
triangles: Lense-Thirring timescales; green squares: apsidal
precession timescales due to the GR correction).  See detailed discussions
in \S~\ref{sec:timescales}.} \label{fig:f1} \end{figure}

\begin{itemize} 
\item
The orbital timescale of a star rotating around the central MBH is 
\begin{eqnarray}
T_{\rm orb}& = &2\pi \sqrt{\frac{a^3}{GM\bh}} \nonumber \\
           & = &1.5 \times 10^3\yr~ m\bh^{-1/2}\left(\frac{a}{0.1\pc}\right)^{1.5},
\label{eq:Torb}
\end{eqnarray}
where $a$ is the orbital semimajor axis of the star.

\item The local two-body relaxation timescale $T_{\rm relax}$ is given by
\begin{eqnarray}
T_{\rm relax}
&\sim &\frac{0.34\sigma^3}{G^2\rho_*\langle m_*\rangle \ln\Lambda}\nonumber \\
&\sim &\frac{2.0\times 10^{9}\yr}{\ln\Lambda} m\bh^{1/2} \frac{M\bh}{M_*(<r)}
       \frac{1\msun}{\langle m_*\rangle} \left(\frac{r}{0.4\pc}\right)^{3/2} \nonumber \\
&\sim &1.2\times 10^9\yr~  m\bh^{1/2} \frac{10}{\ln\Lambda}
       \frac{1\msun}{\langle m_*\rangle} \left(\frac{r}{0.4\pc}\right)^{-0.1},
\end{eqnarray}
where $\langle m_*\rangle$ is the mean stellar mass and $\rho_*$ is the stellar
density (e.g., \citealt{BT87,Alexander99}).  In the estimate of this timescale
below (in Fig.~\ref{fig:f1}), we set $\langle m_*\rangle\sim 1\msun$ and $\ln
\Lambda\sim 10$.

\item The precession due to the stellar cusp: the stellar cusp (mentioned in
\S~\ref{sec:cusp}) may cause stars deviated from purely Keplerian motions and
introduce apsidal precession of their orbits. This apsidal precession timescale
can be estimated by 
\begin{eqnarray}
 T^{\rm cusp}_{\rm prec} 
&\sim & \frac{M\bh}{M_*(<r)}T_{\rm orb}(r) \nonumber \\
&\sim & 7.5\times 10^4\yr~  m\bh^{1/2}
        \left(\frac{r}{0.4\pc}\right)^{-0.1}
\end{eqnarray}
with assuming $M_*(<r)\ll M\bh$.

\item Resonant relaxation timescales (see details in  \citealt{RT96}; 
\citealt{HA06}): in a
timescale much longer than the orbital period and shorter than the apsidal
precession timescale mentioned above, the star can be approximated by a fixed
wire whose mass is the stellar mass, whose shape is a Keplerian ellipse, and
whose linear density is inversely proportional to the local speed in the
elliptical orbit. The wires precess on the timescale $T^{\rm cusp}_{\rm prec}$
and exert mutual torques which induce angular momentum relaxation. The
cumulative effects of the torques result in the change of the absolute value 
of the angular momentum by a timescale of
\be T^{\rm res,S}_{\rm relax}\sim
3.56\times \frac{M\bh+M_*(<r)}{\langle m_*\rangle} T_{\rm orb}(r)
\ee
(hereafter the scalar-resonant relaxation timescale) or the change of the
direction of the angular momentum vector by a timescale of
\be
T^{\rm res,V}_{\rm relax}\sim 0.62\times \frac{M\bh+M_*(<r)}{\langle m_*\rangle}\frac{T_{\rm orb}(r)}{N^{1/2}(<r)}
\ee (hereafter the vector-resonant relaxation timescale), where $N(<r)=M_*(<r)/ \langle m_* \rangle$. 

The sub-parsec massive young stars is a different population from the major
component (old stars with $\langle m_*\rangle=1\msun$ set here) in the Galactic
center.  We still take the above equations as the resonant relaxation
timescales of the young stars, since their number and total mass are much
smaller than the background old stars and the effect of the resonant relaxation
on the young stars should mainly be due to the old stellar population (see the
derivation in \citealt{RT96}).

\item 
The apsidal precession due to the general relativity correction to
the Newtonian equations of motion is given by \citep{Misner73}:
\begin{eqnarray}
T_{\rm GR}
& = &\frac{a c^2 (1-e^2)}{3G M\bh}T_{\rm orb}(a) \nonumber \\ 
& = &3.0\times 10^7\yr~ m\bh^{-3/2} (1-e^2) \left(\frac{a}{0.04\pc}\right)^{5/2},
\label{eq:Pgr}
\end{eqnarray}
where $c$ is the speed of light.

\item 
The Lense-Thirring precession timescale is given by \citep{Misner73}
\begin{eqnarray}
T_{\rm LT}&=&\frac{2\pi a^3(1-e^2)^{3/2}}{2J} \nonumber\\
& = & 2.2\times 10^{10}\yr~ a_*^{-1} m\bh^{-2} (1-e^2)^{3/2} 
      \left(\frac{a}{0.04\pc}\right)^3, \nonumber \\ 
\label{eq:Plt}
\end{eqnarray}
where $J$ is the angular momentum of the MBH and $a_*=J/(G M\bh^2/c)$ is the
dimensionless spin parameter of the MBH.
\end{itemize}

The above timescales are plotted as a function of the distance from
the central MBH in Figure~\ref{fig:f1}, with using the parameters of
the observed stellar cusp described in \S~\ref{sec:cusp} (see a
similar diagram in \citealt{HA06}).  As shown in Figure~\ref{fig:f1},
the timescales $T_{\rm relax}$, $T^{\rm res,S}_{\rm relax}$, and
$T^{\rm res,V}_{\rm relax}$ are all much longer than the stellar age
$\tau_{ \rm age}$ at $r>0.04\pc$, which suggests that the spatial
distribution of the original disk located at $0.04-0.5\pc$ cannot be
erased through either two-body or resonant relaxation processes.  The
apsidal precession timescale $T_{\rm GR}$ is much longer than
$\tau_{\rm age}$ for the stars at $a\ga 0.1\pc$ and it is comparable
to the age of the disk stars with moderate eccentricity at $a\la
0.04\pc$.  The apsidal precession may affect the orbits of the S-stars
(since $T_{\rm GR}<\tau_{\rm age}$ for the S-stars), but it should be
insignificant to the orbital evolution of the disk stars located at $a
\sim 0.04-0.5\pc$.  The Lense-Thirring precession is generally
not important for those disk stars since $T_{\rm LT}\gg \tau_{\rm
age}$ at $a\sim 0.04-0.5\pc$, but it may be important for the
innermost two stars S2 \citep{LB03} and S14 ($T_{\rm LT}\sim 4.1\times
10^6 a_*^{-1}\yr$ for S2 $\sim 8.9\times 10^6 a_*^{-1}\yr$ for S14).
As seen from this Figure, the apsidal precession timescale due to the
stellar cusp is much shorter than the age of the stars but much longer
than their orbital period. This precession affects the secular orbital
evolution of the young stars in the stellar disk.

According to the estimates of these relevant timescales (see
Fig.~\ref{fig:f1}), the disk stars in the GC may be described to be moving in a
stellar cusp with a {\em smooth} potential (the apsidal precession due to the
stellar cusp is naturally included in the description), and the effects of
other precessions, apsidal precession due to the general relativity
correction and the Lense-Thirring precession, may be neglected in the following
calculations.  In this paper, we do not pursue a study on the origin of the
kinematics of the S-stars (with age of a few $10^7\yr$), for which the resonant
relaxation, as well as the precession due to the general relativity correction
and the Lense-Thirring precession, are involved (see also discussions in
\citealt{HA06,Levin07}) and a smooth potential may not be a good approximation
any more.

In addition, if the young stars were formed in a gas disk, we assume that the
gas disk was depleted quickly and gas drag is not important for their orbital
evolution (damping the orbital eccentricity) at least for the two reasons based
on observations: (1) most of the stars were formed in a short timescale $\la
2\Myr$ compared to their age $6\pm2\Myr$; and (2) the MBH is accreting material
via a rate around $10^{-6}-10^{-5} \msun\yr^{-1}$ through a tenuous thick disk
\citep[e.g.,][]{MF01}, which should not have any significant drag on the motion
of the young stars.

\section{Model and numerical experiments}\label{sec:model}

It is plausible that occasionally there may be a star cluster (and/or
an IMBH) in-spiraling into the central region of the Galaxy (see also
discussions in \citealt{PHA07}). In this
section, we study how the orbital configuration of the young stellar
disk are affected by such a perturber. To isolate the problem, we first
estimate how the orbits of the young stars is affected by a perturber
rotating around the MBH at a fixed distance in \S~\ref{subsec:nodal}.
We consider a range of inclination (from 0$\arcdeg$ to
180$\arcdeg$) between the perturber's orbital plane and the stellar
disk plane.  In \S~\ref{subsec:migration}, we illustrate how the
orbital configuration of the stellar disk is affected by the
inward-migration of a perturber.

\subsection{Perturbation on the stellar orbits due to a massive perturber at
a fixed distance from the central MBH}\label{subsec:nodal}

The orbital precession caused by the torque exerted by a massive
perturber at a fixed distance from the central MBH (i.e., on a circular
orbit) changes the inclination angles of the young stars, but not
their eccentricities.  In order for the clockwise rotating young stars
in the inner region to retain their nascent disk's initial
configuration, the oscillation amplitude of the inclination angle,
induced by the nodal precession, needs to be small (say, $<20\arcdeg$;
this number is roughly the maximum range of the estimated inclination
angles of the young stars in the inner disk relative to the disk
normal; \citealt{Bel06}). The oscillation amplitude should be small if the
inclination angle between the orbital plane of the massive perturber and the
stellar disk, denoted by $\beta$, is close to 0\arcdeg or 180\arcdeg, but
it can be large for other intermediate values of $\beta$ unless the
precession timescale is much longer than the stellar age.

\begin{figure} \epsscale{1.0}
\plotone{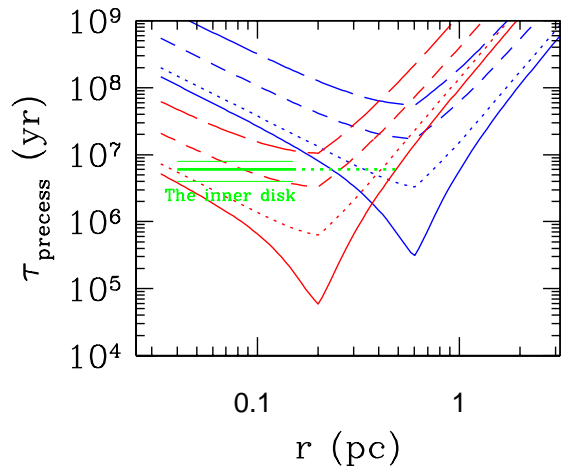}
\caption{The timescales of the nodal precession of stellar orbits due to the
torque exerted by a massive perturber with mass $M\p=10^4\msun$ at a fixed
distance from the central MBH $r\p$. The stars are on nearly circular orbits
with the Galactocentric distance $r$. The left (red) and right (blue) sets of lines are for
a perturber located at $r\p=0.2\pc$ and $r\p=0.6\pc$, respectively. For each
set, different line types represent different inclination angles $\beta$
between the orbital plane of the massive perturber and that of the stellar disk
[from top to bottom: $\beta=85\arcdeg$ (long-dashed lines), 75\arcdeg
(short-dashed lines), 45\arcdeg (dotted lines), and 15\arcdeg (solid lines)].
The age and position of the inner stellar disk ($\tau_{\rm age}\sim 6\pm2\Myr$
and $r\sim$ 0.04--0.15\pc) observed in the GC are marked as solid green lines.
The dotted green line illustrates the extension of the assumed original disk to
$\sim 0.5\pc$. To preserve the coplanar structure of the stars in the region
0.04--0.15$\pc$ not to be significantly affected by the nodal precession, the
perturber should at least (i) be located at a far distance (e.g.,
$r\p\sim0.6\pc$), or (ii) have an orbit nearly parallel or perpendicular to the
disk plane if the perturber is located at a small distance (e.g., $r\p\sim0.2\pc$),
or (iii) have a substantially smaller mass. The orbits of the stars at the
outer region ($r\sim$0.15--0.5\pc) can generally be affected by the nodal 
precession at some values of $\beta$ unless the perturber mass is substantially
smaller.  See detailed discussion in \S~\ref{subsec:nodal}.
} \label{fig:f2} \end{figure}

Approximating the gravity of the perturber and the stars as rings, their nodal
precession frequency is estimated to be \citep{Nayakshin05,Nayakshin06}
\be 
\omega_{*}=\frac{G M\p}{(r^2+r\p^2)^{3/2}}\frac{r\p}{r}
           \frac{1}{\Omega_{\rm K}} I(\delta,\beta),
\label{eq:domegadt}
\ee
where $M\p$ is the mass of the perturber which is much more massive than a
young star, $\Omega_{\rm K}$ is the Keplerian angular frequency of the stars,
$\delta=\frac{2r\p r}{r\p^2+r^2}$, 
$I(\delta,\beta)=\int^{2\pi}_0 \frac{d\phi}{2\pi}\int^{2\pi}_0
\frac{d\phi'}{2\pi}\frac{\sin\phi'\sin\phi}
{[1-\delta(\cos\beta\sin\phi'\sin\phi+\cos\phi'\cos\phi)]^{3/2}}$, and
$I(\delta,\beta)=I(\delta,\pi-\beta)$.  
Using equation (\ref{eq:domegadt}), we show the precession timescale
$\tau_{\rm precess}\equiv 1/\omega_{*}\propto 1/M\p$ in
Figure~\ref{fig:f2} with different parameters $r\p$ and $\beta$. As
seen from Figure~\ref{fig:f2}, for a perturber with mass $M\p=10^4\msun$
and located at a relatively large Galactocentric distance ($r\p\sim0.6\pc$; 
blue lines), the precession timescales in some region ($r\sim 0.5\pc$) 
may be smaller than the stellar age for some values of $\beta$ and the 
orbital inclinations of the stars in this region may have substantial 
variations due to the nodal precession; but the precession timescale at 
$r\sim0.04-0.15\pc$ is generally much longer than the stellar age and 
the initial coplanar structure of the inner disk can be well preserved 
for any values of $\beta$. If the perturber is located at a smaller 
Galactocentric distance ($r\p\sim 0.2\pc$; red lines), the nodal precession
timescale of the young stars at $r\sim0.1\pc$ is much shorter than
or comparable with their age if $\beta$ is in the range
$\sim$10\arcdeg--80\arcdeg (or $\sim$100\arcdeg--170\arcdeg);
thus the orbital inclinations of the stars not only in the outer region
$\sim$0.15--0.5$\pc$ but also in the inner disk may have substantial variations
or a warpness may be developed in the stellar disk due to the nodal precession
(these qualitative estimates are confirmed by direct numerical calculations
of the stellar orbital motion in \S~\ref{subsec:migration}).
If the perturber's orbit is almost perpendicular to the initial disk
plane (red long-dashed line), the precession timescale for the young
stars in the inner region is long enough compared to their age, and thus
the inner disk may be not affected by the precession and its initial coplanar
structure can be preserved. If the perturber orbit is nearly parallel to the
disk plane, as mentioned in the paragraph above, although Figure~\ref{fig:f2} shows that
the precession timescale is short compared to the stellar age, the disk
configuration in the inner region may still be preserved since the
oscillation of the normal to the stellar orbital planes around the normal
to the perturber orbital plane is small. Thus, the preservation of the inner
disk suggests that if the perturber with mass $10^4\msun$
is located at $r\p\sim 0.2\pc$, its orbit should be more likely to be parallel
or perpendicular to the assumed primary disk.  This result is also
demonstrated in \S~\ref{subsec:migration} below.  
For a substantially smaller $M\p$, the precession timescale $\tau_{\rm
precess}$ can be long enough so that the inner disk may be maintained for a
large range of $\beta$.

\subsection{Perturbation on the stellar orbits by an inward-migrating 
perturber}\label{subsec:migration}

In this subsection, we study how the background stellar orbits evolve
as a massive nearby perturber migrating inward.  For simplicity, we
assume the perturber to be an IMBH with a point mass potential.  If
the perturber is a dark cluster, the gradually intensifying Galactic
tidal effect is likely to induce the perturber to lose mass during its
inward migration.  We also consider the possibility that the perturber
has a declining mass.

Before the perturber becomes bound to the central MBH, the perturber spirals
into the Galactic center under the action of dynamical friction which induces
a deceleration 
\be
\frac{d\vec{v}\p}{dt}=-\frac{\vec{v}\p}{\tau}-\nabla\Phi(r),
\label{eq:dvpdt}
\ee
where the dynamical-friction timescale
\be
\tau=t\df=\frac{v\p^3}{8\pi G^2\ln\Lambda M\p \rho(r)
          [{\rm erf}(X)-\frac{2X}{\sqrt{\pi}}exp(-X^2)]},
\label{eq:tdf}
\ee $v\p=|\vec{v}\p|$ is the velocity of the perturber,
$X=\frac{v\p}{\sqrt{2}\sigma(r)}$, and $\ln\Lambda$ is the logarithm
of the ratio of the maximum and minimum impact parameters and
$\Lambda\simeq M\BH/M\p$ \citep{BT87}.  As the perturber or the IMBH
migrates inward and forms a bound binary black hole (BBH) with the
central MBH at $r\p\simeq a\H$ (see eq.~\ref{eq:aH}), it continues to
lose energy and angular momentum through dynamical friction.  But, the
influence of dynamical friction on the IMBH's orbit becomes less
efficient as its orbital period decreases and its orbital velocity
increases. After the BBH becomes hard at $a\h=(M\p/4M\BH)a\H \simeq
0.004\pc~m\bh \left(\frac{M\p/M\bh}{0.01}\right)
\left(\frac{100\kms}{\sigma}\right)^2$, it
loses energy mainly through three-body interactions with stars passing
by its vicinity.  The orbital decay timescale of a hard BBH in the GC
is about $t\h\equiv r\p/\dot{r}\p\simeq 6\times 10^9\yr$ (see eq.~38
in \citealt{YT03}).  The gravitational radiation timescale of the BBH
is longer than the Hubble time (see eq.~39 in \citealt{YT03}) and it
is unlikely to be significant in the spatial range of the BBH
considered in this paper.  During the transition stage (after the BBH
becomes bound but before it becomes hard), the BBH's hardening
timescale may be higher than the estimate from the dynamical friction
timescale $t\df$($\sim 10^6\yr$ at $1\pc$ for $M\p\sim 10^4\msun$), as
this process becomes less efficient, and it increases to that at the
hard stage $t\h$ as the IMBH migrates in \citep{Y02}.  
Under the constraint placed
on the distance between Sgr A$^*$ and the center of mass of the BBH
(eq.~35 or Fig.~2 in \citealt{YT03}), the mass of the secondary BH (or
star cluster)
$M\p$ should be smaller than $\sim 0.03M\bh$ if it is located at
$0.3\pc$ and smaller than $\sim 0.08M\bh$ if it is located at
$0.1\pc$.  If the GC (or the central MBH) was formed through the
assemblage of sinking star clusters (or IMBHs, which are considered to
be candidates for the massive perturber in this paper), we can also
infer their masses on the assumption that we are not living in a
unique time. Such an argument would lead to a mass estimate for a
typical perturber to be $ \sim [M\bh+M_*(<1\pc)]\times 10^7\yr/10^{10}
\yr \sim 5\times 10^3\msun$.  Finally, star clusters with mass around
$10^4\msun$ are found within several tens $\pc$ from the GC.  Based
on these considerations, we adopt $M\p= 10^4\msun$ as a fiducial mass
for the massive perturber in general.  We also consider the potential
implications for other values of $M\p$.

\begin{figure} \epsscale{1.0}
\plotone{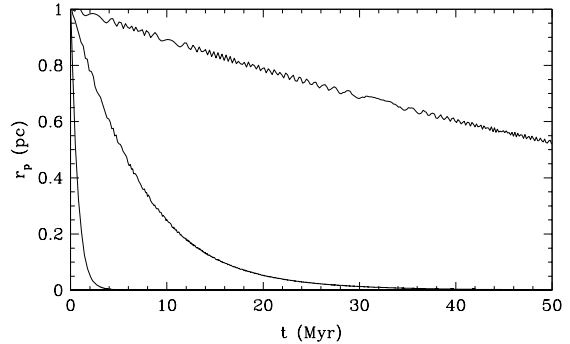}
\caption{The orbital evolution of a perturber used in the dynamical
model proposed in \S~\ref{subsec:migration}. The perturber is initially
on a nearly circular orbit at $1\pc$ and its evolution follows equation
(\ref{eq:dvpdt}) with $\tau=10^8,10^7,10^6\yr$ from top to bottom.}
\label{fig:rpt} \end{figure}

In the first set of simulations of stellar responses to a migrating
perturber, its initial orbit is assumed to be nearly circular with a
radius $1\pc$.  The path of inward migration follows
equation (\ref{eq:tdf}), where the migration timescale $\tau$ is set
to several different values ($10^6, 10^7, 10^8\yr$).
The initial conditions of the perturber are set so that it attains an
(almost constant) eccentricity $\sim 0.02$ after it moves sufficiently close
to the MBH ($r\p\la0.2\pc$).
(In subsequent models, the potential implications for the higher values of
the perturber's eccentricity are also considered. But we do not consider
a highly eccentric orbit for the perturber, since the
orbital decay of a dark cluster or a secondary BH via dynamical friction and
three-body interactions with the central MBH and stars generally does not
result in a highly eccentric orbit, e.g., with eccentricity $\la 0.3$;
\citealt{PR94,Q96}.)  The orbital
evolution of the perturber is shown in Figure \ref{fig:rpt}.  These
results indicate that $15\Myr$ (or $10\Myr$) would be required for the
perturber to migrate inward from $1\pc$ (or $0.5\pc$) to $0.1\pc$ if
$\tau=10^7\yr$.  In this paper, we do not address the interesting
issue whether the migration of the perturber may be correlated with
the formation of the young stars.  We simply assume that the young
stars formed throughout the disk prior to or during the migration of
the perturber.  Note that the time $t=0$ does not necessarily
represent the time when the young stars formed.

We simulate the orbital evolution of 75 test particles
as young stars, with the logarithm of their initial semimajor axes $a$
distributed uniformly in the radial range $\lg(0.04\pc)$--$\lg(0.5\pc)$.
Each star moves independently in the potential of the central MBH, the
inward-migrating IMBH, and the stellar cusp.
We set the inclination of the
perturber to zero and it does not change in the spherical
gravitational potential of the cusp.  Five sets of initial inclination
angles of young stars $i$ are chosen randomly in the range
0\arcdeg--10\arcdeg, 20\arcdeg--30\arcdeg, 80\arcdeg--90\arcdeg,
110\arcdeg--120\arcdeg, and
170\arcdeg--180\arcdeg, respectively.  Their other orbital elements
relative to the central MBH, such as the longitude of ascending node,
argument of pericenter, and true anomaly, are chosen randomly in the
range $[0\arcdeg,360\arcdeg]$.  The initial eccentricities $e$'s of the orbits
around the MBH are set to zero.  But, they rapidly attain finite
amplitudes, due to the gravity contributed by the stellar cusp.  
Given the position and velocity of a test particle at a given time, we can
still define its time(or position)-dependent semimajor axis and
eccentricity by assuming that the particle is on an elliptical orbit
around the MBH at the appropriate time.  In order to illustrate the
change of the orbital configuration of a test particle, the angle of
the orbital plane of each particle relative to its initial plane,
$\theta$, is used below, which may describe the observed inclination
angles of the young stars relative to the observed inner disk
(our conclusions do not change if the initial longitudes of ascending
node of the particles in the simulations are set to be the same so that
all the particles are initially on a thin disk).
During their evolution, the orbits of two test particles with the same
inclination angles do not generally lie on the same plane.

\begin{figure} \epsscale{1.0}
\plotone{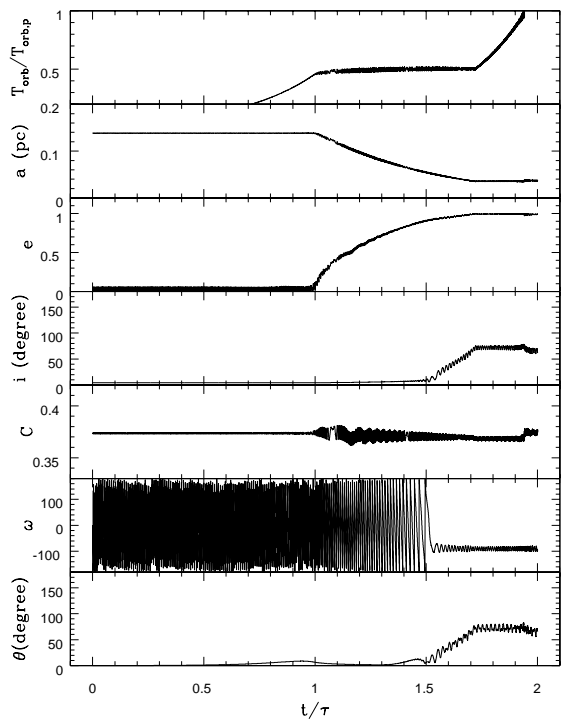}
\caption{The orbital evolution of a test particle in the dynamical
model proposed in \S~\ref{subsec:migration}. The initial inclination
angle of the particle is close to 0. From top to bottom, the panels
show orbital period ratio to the perturber, semimajor axis,
eccentricity, inclination angle to the orbital plane of the perturber,
the quantity $C$ defined in \S~\ref{subsec:migration}, argument of
pericenter, and the angle between the orbital plane of the particle
and its initial orbital plane $\theta$. The particle is first captured
into the 2:1 mean-motion resonance and then released from the
resonance but into the $\omega=\pm 90\arcdeg$ secular resonance.}
\label{fig:reson1} \end{figure}

\begin{figure} \epsscale{1.0}
\plotone{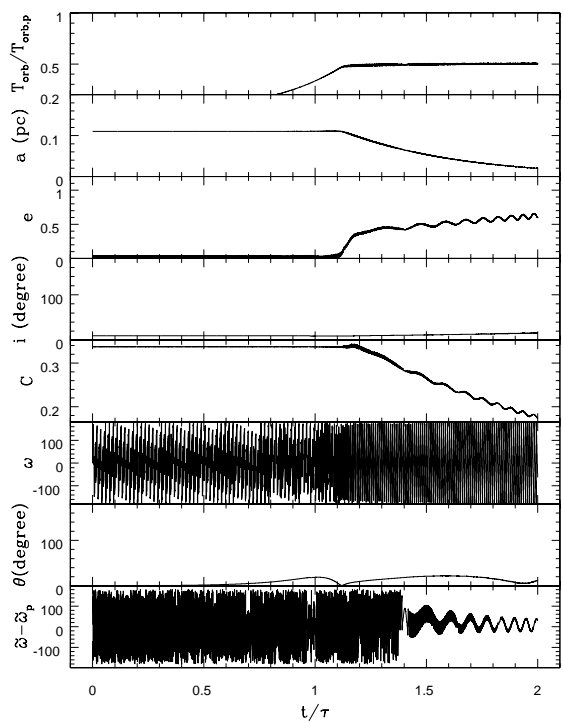}
\caption{The orbital evolution of a test particle that is captured
into the 2:1 mean-motion resonance and then into
$\bar\omega-\bar\omega\p=0\arcdeg$ secular resonance, where
$\bar\omega-\bar\omega\p$ is the longitude difference of pericenter
between the particle and the perturber.}
\label{fig:reson2} \end{figure}

\begin{figure} \epsscale{1.0}
\plotone{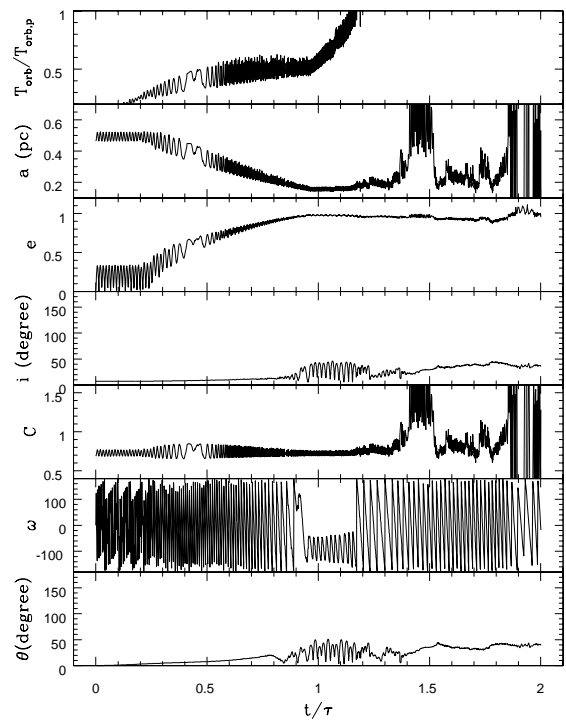}
\caption{The orbital evolution of a test particle that initially
follows the evolutionary pattern as that in Figure~\ref{fig:reson1}
and then is on an irregular orbit.} \label{fig:irregular} \end{figure}

We first show the simulation results for a model in which the
perturber lies almost on the initial orbital plane of the disk stars
($i\in [0\arcdeg,10\arcdeg]$). We set the perturber mass to be
$M\p=10^4\msun$ and the migration timescale to be $\tau=10^7\yr$. From
the simulation results, we find a range of stellar responses.  For
illustration purpose, it is useful to classify these responses into
the following categories with representative cases shown in
Figures~\ref{fig:reson1}--\ref{fig:irregular}.

\begin{figure*} \epsscale{0.9}
\plotone{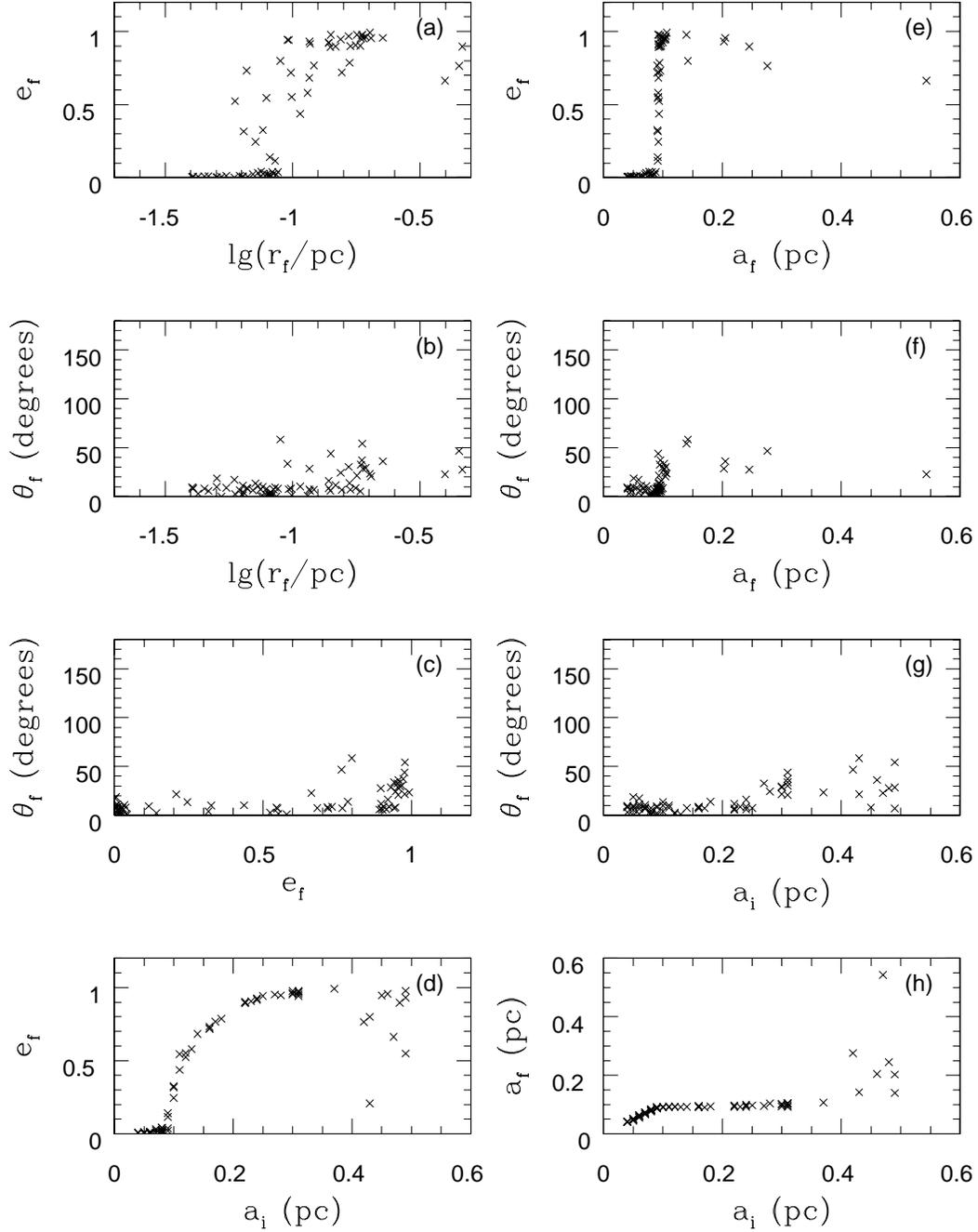}
\caption{Orbital distribution of simulated test particles when the
perturber with mass $10^4\msun$ migrates to $0.15\pc$ (denoted by the subscript ``f''; the initial
values of their orbital parameters are denoted by the subscript ``i'').
The initial inclination angles of the
particles to the perturber plane are in the range
$[0\arcdeg,10\arcdeg]$.  Note that the distance to the central MBH
$r_f$ is generally different from the semimajor axis of the particle
$a_f$ due to the non-zero eccentricity $e_f$. See \S~\ref{subsec:migration}.}
\label{fig:distr} \end{figure*}
                                                                                                           
\begin{figure*} \epsscale{0.9}
\plotone{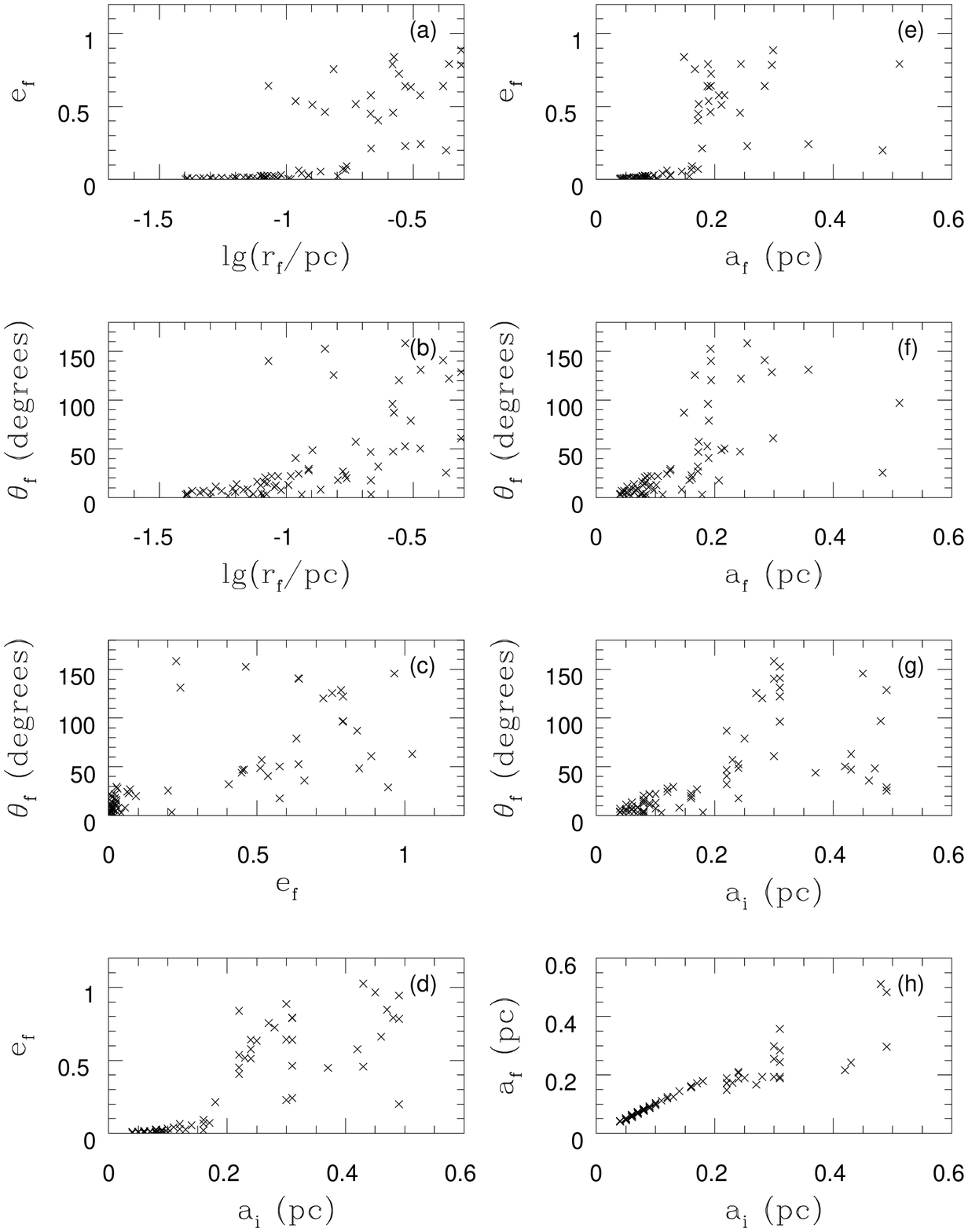}
\caption{Orbital distribution of simulated test particles with initial
inclination angles $i\in[80\arcdeg,90\arcdeg]$, when the perturber with
mass $M\p=3\times 10^4\msun$ migrates to $r\p=0.3\pc$. Other initial
conditions are the same as those in Figure~\ref{fig:distr}.}
\label{fig:distrinc90} \end{figure*}

A) Figure \ref{fig:reson1} shows the evolution of a test particle
($i\in[0\arcdeg,10\arcdeg]$) which is captured into the 2:1
mean-motion resonance of inwardly migrating perturber, i.e., their
orbital period ratio becomes 2:1 at $t/\tau\simeq 1$. After being
captured, the particle migrates inward with the decaying perturber.
During the resonant march, the particle retains an adiabatic invariant
which for a $(p+q,p)$ mean-motion resonance is $C\equiv
a^{1/2}[(p+q)-p(1-e^2)^{1/2}\cos i]$ \citep{YT01}.  The magnitude of
$p>0$ and $q>-p$ are integers, and $p=q=1$ for the inner (2,1)
resonance (by ``inner'' refers to the resonances lie inside the
perturber's orbit, i.e. with $a<a\p$). For this adiabatic invariant, a
reduction in the particle's semimajor axis during the migration can
be compensated by an increase in its $e$ and $i$.  The results in
Figure \ref{fig:reson1} indeed indicate a growth in the particle's
eccentricity.  When its eccentricity is sufficiently high ($\sim 0.9$
at $t/\tau\simeq 1.5$), its inclination also grows.  A comparison
between the results of the two sets of simulations indicates that the
inclusion of the stellar cusp in the gravitational potential promotes
the elevation of the particle's inclination to a range between
0\arcdeg--90\arcdeg, rather than to a polar ($i=90\arcdeg$) or a
retrograde orbits ($i=180\arcdeg$) around a point-mass potential
\citep{YT01}. The asymptotic inclination decreases with increasing
fractional contribution of the stellar cusp mass to the gravitational
potential.  During the lifting of its inclination, the particle
enters into a secular resonance with perturber's argument of the
pericenter $\omega$, librating around $\pm 90\arcdeg$.  At an advanced
stage of evolution ($t/\tau \sim 1.7$), the particle is released from
the perturber's mean-motion resonance when it attains relatively large
eccentricity and inclination.  But, the particle retains its secular
resonance ($\omega\sim \pm90\arcdeg$) with the perturber which
prevents close encounters between them.  Consequently, the orbit of
the particle is stable on the timescale of our integration ($\sim
5\times 10^2$ initial orbital periods of the perturber).

B) Figure \ref{fig:reson2} illustrates the evolution of second
representative particle with a similar $i\in[0\arcdeg,10\arcdeg]$ but
have a different initial location to that in the previous example.
The initial evolution of this particle is generally the same as that
in the previous model (Figure \ref{fig:reson1}).  
The particle
is captured into the perturber's mean-motion resonance of the
perturber at $t/\tau\ga1.1$ (see Figure \ref{fig:reson2}).  At
$t/\tau\sim1.4$, the test particle enters into a state of secular
resonance with the perturber such that the difference of their
pericenter longitude $\eta\equiv\bar\omega-\bar\omega\p$ (where ${\bar
\omega}= \Omega + \omega$ and $\Omega$ is the longitude of ascending
node) librating around 0.  During this phase, the quantity $C$ is no
longer an invariant. The capture of the particle into this perturber's
secular resonance can be understood through the evolution of the
difference between the precession rates of the perturber and the test
particle, which may be described by $d\eta/dt=A_1+A_2\cos(\eta)$
\citep{NLI03}.  The right-hand side of this evolution equation
includes the contribution from the precession of the test particle
induced by the secular interaction with the perturber and also the
contribution from the precession of both the perturber and the test
particle induced by the stellar cusp.  As the perturber and the test
particle migrate inward, both of these contributions evolve with time.
At some point during the course of migration, $A_1\ll A_2$ and the
longitude difference of pericenter librates around $\eta=0$ (or
180\arcdeg). The eccentricity ratio of the test particle and the
perturber $e/e\p$ [which evolves as $d(e/e\p)/dt\propto \sin\eta$;
\citealt{NLI03}] also librates around an equilibrium value.  As the
test particle continues to migrate inward with the perturber, it may
retain the state of secular resonance but the equilibrium value of the
eccentricity ratio changes as the Hamiltonian of the system evolves.
Similar secular resonance was discussed in \citet{NLI03}, where the
depletion of a protostellar disk around a planetary system causes the
change of the precession rates of the system. In this paper, the
change of the precession rates is caused by the migration of the
system in a stellar cusp.

C) Figure \ref{fig:irregular} shows a third example of the evolution
of a test particle ($i\in[0\arcdeg,10\arcdeg]$), which has a similar
evolutionary pattern as that in Figure \ref{fig:reson1} at $t/\tau\la
1.4$ and is on an irregular orbit afterwards.

The above three categories essentially convert the entire domain of
stellar responses.  But the actual outcome of the perturber's passage
sensitively depends on the particles' initial orbital parameters.
In this set of simulations, most of the particles with initial semimajor
axis in the range 0.2--0.5$\pc$ fall into case C, and the stars with
smaller semimajor axes generally fall into case A and a few in case B.
Figure \ref{fig:distr} shows the distribution of the orbital
parameters of the test particles.  In this figure, the subscripts
``i'' and ``f'' of various variables represent their initial and final
values, respectively. For the ``final'' values, we refer to those when
the perturber migrates to $r\p=0.15\pc$ ($t/\tau\simeq1.1$). (Note that
the value of $r\p$ is an instantaneous position of the perturber, and
the perturber does not necessarily stop there but continues to migrate inward
under the action of dynamical friction with time going on.) As seen
from Figure \ref{fig:distr}(g), (d), and (h), the semimajor axes,
eccentricities, and inclination angles of particles with small initial
semimajor axis ($a_i\la 0.1\pc$) are not greatly affected by the
migration of the perturber.  Test particles with initial semimajor
axes in the range 0.1--0.3$\pc$ have essentially the same final
semimajor axes ($\sim 0.1\pc$) because these particles are captured
into the 2:1 mean-motion resonance and forced to migrate with the
perturber during its orbital evolution.  As seen from panels (d)--(g),
their eccentricities and inclinations may be excited to high values.
Panel (c) indicates that the test particles with high inclination
angles generally have high eccentricities.  The particles with the
same semimajor axes may have different instantaneous distances $r_f$
from the central MBH due to the non-zero eccentricity $e_f$.  Panels
(a) and (b) show the distribution of the eccentricities and
inclination angles versus the distances.  Panel (a) indicates the
tendency that the eccentricities of the test particles are low at the
small radii and high at the large radii.  Panel (b) indicates that the
stellar disk may be maintained at the inner radii [$\lg(r_f/\pc)\la
-0.8$] and the test particles at the outer radii have high inclination
angles relative to the inner disk.  Most particles with initial
semimajor axes larger than $\sim0.35\pc$ are captured into 2:1
resonance during the early stages of the perturber's evolution but
they are released from the resonance and attained irregular orbits
after the perturber has migrated inside $r\p=0.15\pc$.  

We also check the dependence of the final orbital distribution on the
perturber's position $r\p$, mass $M\p$, and eccentricity. For a smaller
$r\p=0.1\pc$, the inclination angles and eccentricities of the stars at the
inner radii ($r_f<0.1\pc$) may also be excited to high values and the inner
disk can no longer be maintained. For a larger $r\p=0.3\pc$, although the
`final' eccentricity distribution is qualitatively similar to those shown in
Figure~\ref{fig:distr}, but the inclination angles $\theta_f$ cannot be excited
beyond $20\arcdeg$. 
A relatively low-mass perturber (e.g., $M\p=10^3\msun$)
cannot significantly excite the inclination angles, either, although it can
excite some particles' eccentricities to high values by capturing them into the
2:1 or 3:2 mean-motion resonance.  An increase in the perturber mass (e.g.,
$M\p=3\times 10^4\msun$) enhances its secular perturbation and widens its
mean-motion resonances.  So does an increase in the perturber's eccentricity or
an increase in the test particles' initial eccentricities of the test particles.
In these cases, some particles are captured into
mean-motion resonances other than the 2:1 resonance, such as, the 3:2, 3:1, or
4:1 resonance. Some particles may transit from one mean-motion resonance to
another mean-motion resonance during their orbital evolution.  Relatively more
particles are captured into secular resonance
$\bar\omega-\bar\omega\p=0\arcdeg$ or 180\arcdeg. 

\begin{figure} \epsscale{1.0}
\plotone{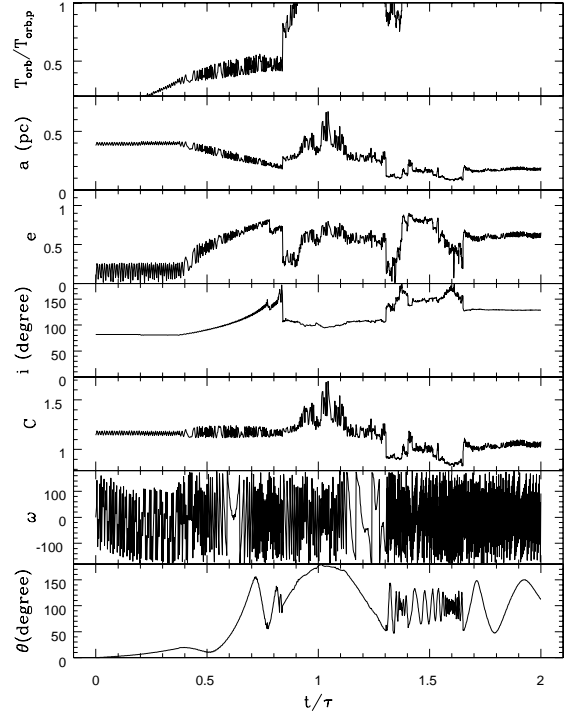}
\caption{The orbital evolution of a test particle with initial
inclination angle close to $90\arcdeg$. The particle is first captured
into the 2:1 mean-motion resonance and then is on an irregular
orbit. }
\label{fig:reson3i90} \end{figure}
                                                                                                           
\begin{figure} \epsscale{1.0}
\plotone{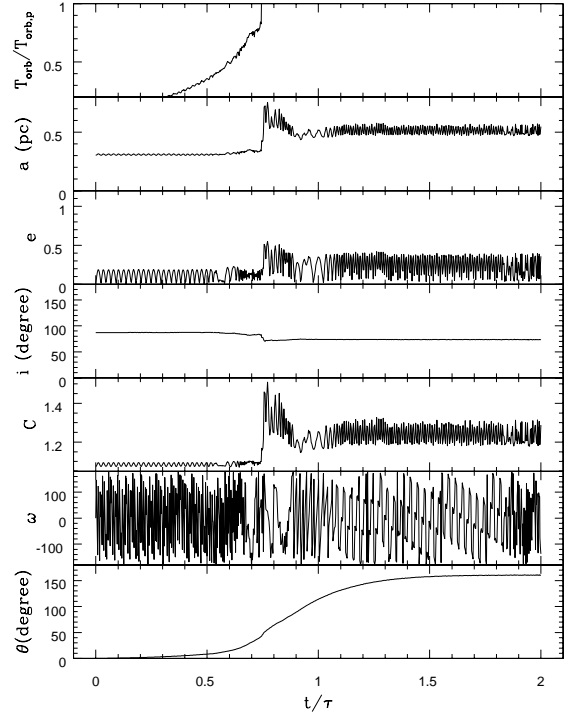}
\caption{The orbital evolution of a test particle with initial
inclination angle close to $90\arcdeg$. Although the inclination angle
of the particle relative to the perturber orbital plane $i$ changes
little, the angle of the particle orbital plane to its initial orbital
plane $\theta$ changes significantly due to the nodal precession in
the gravitational potential of the stellar cusp and the change of the
inclination angle from an initial value close to $90\arcdeg$ to $\sim
70\arcdeg$ at $t\sim 0.75\tau$ caused by a close encounter.
The angle $\theta$ is almost constant or does not oscillate after
$t/\tau\sim 1.5$
because the perturber has migrated into the inner region and
does not have significant effects on the orbital evolution of the outside
particle anymore.
}
\label{fig:reson3i90p} \end{figure}

\begin{figure*} \epsscale{0.9}
\plotone{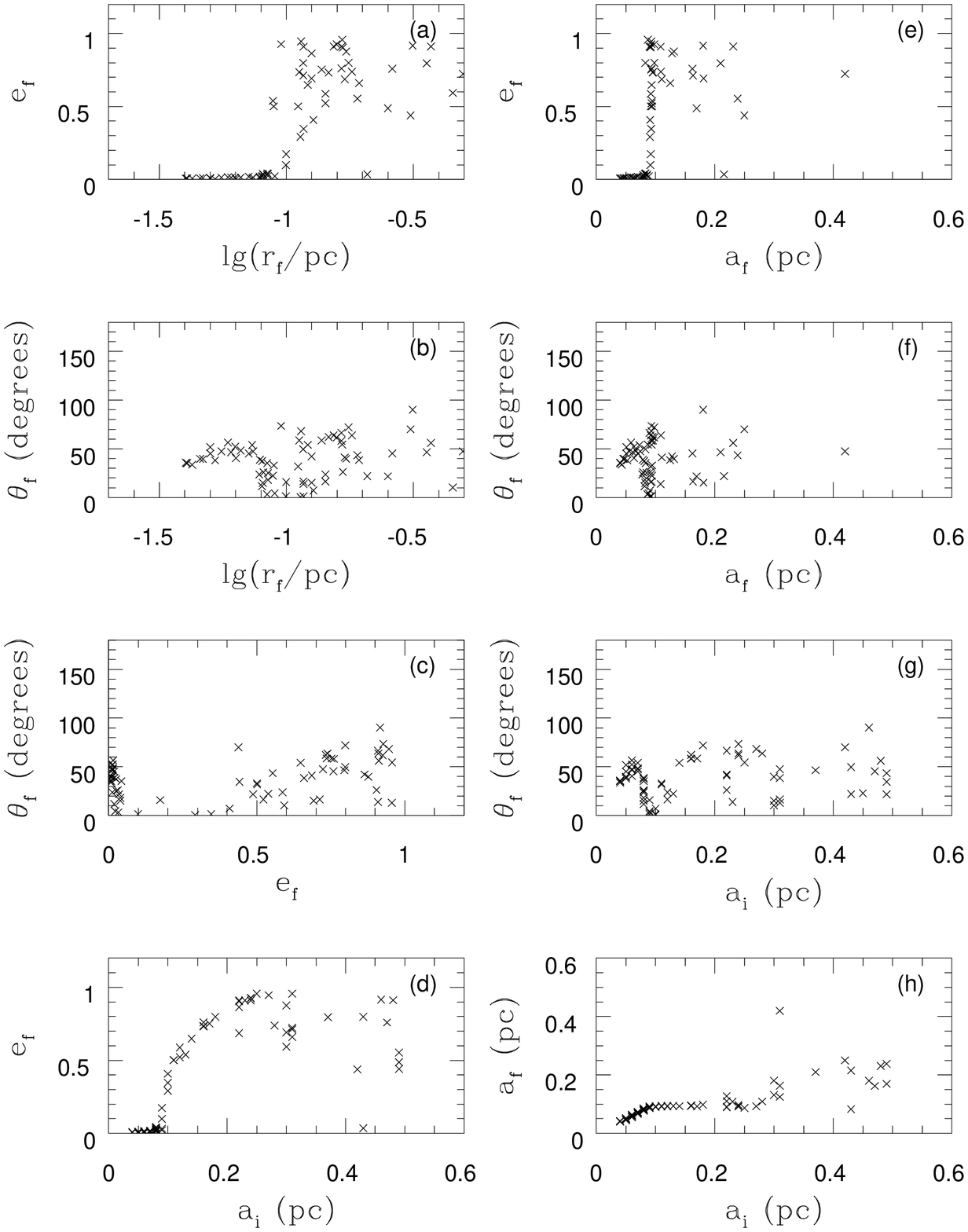}
\caption{Orbital distribution of simulated test particles with initial
inclination angles $i\in[20\arcdeg,30\arcdeg]$ when the perturber with
mass $10^4\msun$ migrates to 0.15\pc. Other initial
conditions are the same as those in Figure~\ref{fig:distr}.}
\label{fig:distrinc30}
\end{figure*}
                                                                                                           
\begin{figure*} \epsscale{0.8}
\plotone{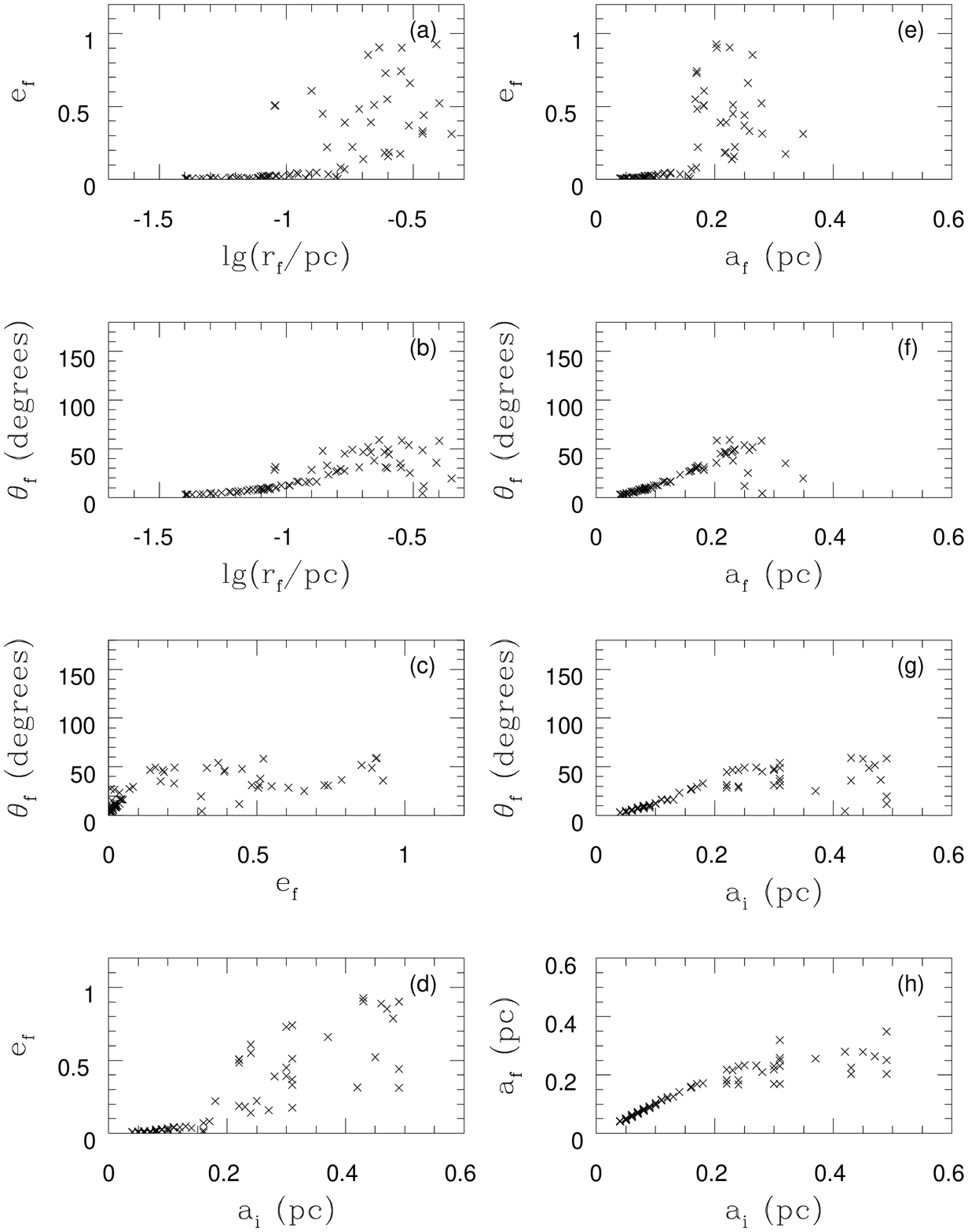}
\caption{Orbital distribution of simulated test particles with initial
inclination angles $i\in[20\arcdeg,30\arcdeg]$ when the perturber with
mass $5\times10^3\msun$ migrates to 0.3\pc. Other initial
conditions are the same as those in Figure~\ref{fig:distr}.}
\label{fig:distrinc30rp0.3}
\end{figure*}

We now consider the possibility that the perturber migrates to the
proximity of the stellar disk on a highly inclined orbit.  We show the
case that the initial disk plane is almost perpendicular to the
perturber orbital plane in Figure \ref{fig:distrinc90}, for which
$M\p=3\times 10^4\msun$ and $r\p=0.3\pc$.  In this case, most of the
main features of the distribution in Figure~\ref{fig:distr} are
preserved, including the maintenance of a coherent inner stellar disk
with $\theta_f\la 20\arcdeg$. The angles of the final orbital plane of
the stars relative to their initial disk plane $\theta_f$ have a much
larger range 0\arcdeg--180$\arcdeg$ than those shown in
Figure~\ref{fig:distr}.  Figures~\ref{fig:reson3i90} and
\ref{fig:reson3i90p} illustrate two representative particles' orbital
evolution.  In Figure~\ref{fig:reson3i90}, after the particle is
captured into the perturber's 2:1 mean-motion resonance (or 3:2 resonance
for some other particles), both its
eccentricity and its inclination also increase as the particle
migrates inward.  The magnitude of $C$ is an adiabatic invariant at
$t/\tau\la 0.8$ and the particle is on an irregular orbit after it
undergoes a close encounter with the perturber at that epoch.  In
Figure~\ref{fig:reson3i90p}, the inclination angle of this particle
$i$ changes little during the evolution; but after $i$ flips from an
initial value close to $90\arcdeg$ to $\sim 70\arcdeg$ because of its
close encounter with the perturber at $t/\tau\sim$0.7--0.8, the angle
between the orbital plane of the particle and its initial orbital
plane changes significantly due to the nodal precession of the
particle orbital plane.  The excitation of the eccentricities
of the test particles shown in Figure \ref{fig:distrinc90} results
from their capture into mean-motion resonances and close encounters with
the perturber. In addition to resonance capture and close encounters,
nodal precession due to the perturber also results in the excitation of
the inclination angles $\theta$.
The distribution of the final semimajor axes of the particles
in the outer region (e.g., see $a_f>0.15\pc$ in Fig.~\ref{fig:distrinc90}e)
is broader than that shown in Figure \ref{fig:distr} due to the scattering
through close encounters with the perturber and wider resonances that they
are captured into.

We also calculate the orbital distributions of the particles if their initial
inclination angles to the perturber's orbital plane $i$ are in the range
20\arcdeg--30\arcdeg, 110\arcdeg--120\arcdeg, and 170\arcdeg--180\arcdeg,
respectively.  For the cases of initial $i\in[20\arcdeg,30\arcdeg]$ and
$i\in[110\arcdeg,120\arcdeg]$, if the perturber migrates to a position
$r\p=0.15\pc$, the stars in the inner radii cannot be maintained on a disk
(with $\theta_f$ up to $50\arcdeg$ or 120\arcdeg, e.g., see
Fig.~\ref{fig:distrinc30}) due to the nodal precession of the particles'
orbital planes, which is consistent with our discussion in
\S~\ref{subsec:nodal}.  If the perturber has a smaller mass and migrates to a
relatively outer position (e.g., $M\p=5\times10^3\msun$ and $r\p=0.3\pc$), the
angles $\theta_f$ of the inner stars can be small (e.g., $\la20\arcdeg$), but a
warpness may be developed in the disk (the warpness is not indicated by
observations so far).  For the case of initial $i\in[20\arcdeg,30\arcdeg]$ (see
Fig.~\ref{fig:distrinc30rp0.3}), both the eccentricities and inclinations of
the stars at the outer region can be excited up due to capture into mean-motion
resonances, close encounters with the perturber, and nodal precessions. But the
eccentricities cannot be sufficiently excited up in the case of initial
$i\in[110\arcdeg,120\arcdeg]$ (e.g., with $e_f\la 0.6$), because our
simulations show that they are less likely to be captured into mean-motion
resonances compared to the case of $i\in[20\arcdeg,30\arcdeg]$ or that their
mean-motion resonances are more likely to be unstable and affected by close
encounters with the perturber. A relatively higher perturber mass may increase
the excited eccentricities due to close encounters, but it also induces larger
nodal precession of stars at the inner radii.

For the case of initial $i\in[170\arcdeg,180\arcdeg]$, the inner stars may be
maintained on a disk and on nearly circular orbits.  The angles $\theta_f$ of
the stars with large initial semimajor axes (e.g., $a_i>0.3\pc$) can be excited
up to $50\arcdeg$, and their eccentricities can also be excited up.  But many
excited particles have large $a_f$ (e.g., $\ga 0.5\pc$) and the excited
eccentricity $e_f$ of the particles at $r\sim 0.15-0.5\pc$ are not high enough
(e.g., $\la 0.6$) compared to the observations shown in Figure \ref{fig:f3}.
In this case, both of the inclination and eccentricity excitations come mainly
from close encounters with the perturber.

If the perturber is a dark cluster (with a central IMBH), its orbital decay may
be accompanied with mass loss induced by the strong external tidal potential.
Our simulations indicate that for the same initial perturber mass, a decrease
in the perturber mass during its orbital decay may make the inner stellar disk
more likely to be preserved.

We also test the effects of different perturber-migration
timescales. For longer migration timescales (e.g., $5\Myr$ or
$10\Myr$), the main results are generally not affected. Note that a
very large ($\gg 10\Myr$) migration timescale is not compatible
with the estimated age of the young stars.  For a substantially
shorter migration timescale (e.g., $\tau=1\Myr$), the inward-migration
of the perturber is too fast for some particles to be captured into
the mean-motion resonance and the excitation of the eccentricity and
inclination becomes insufficient.  

\subsection{Comparison with observations and discussions} 
\label{subsec:obs} 
 
The numerical models in \S~\ref{subsec:migration} simulate the
interaction between a disk of stars with nearly circular orbits and an
inward-migrating perturber. Provided the perturber's orbit is either
nearly coplanar (in corotating directions)
or essentially overhead, the young stars at the outer region generally
attain high eccentricities (up to unity), while those at the inner region
retain low
eccentricities.  In addition, the young stars at the outer region tend
to have a large range of inclination angle relative to the disk, while
those in the inner region roughly remain in the initial disk (see
panels a and b in Figs.~\ref{fig:distr} and \ref{fig:distrinc90}).

\begin{figure} \epsscale{1.0}
\plotone{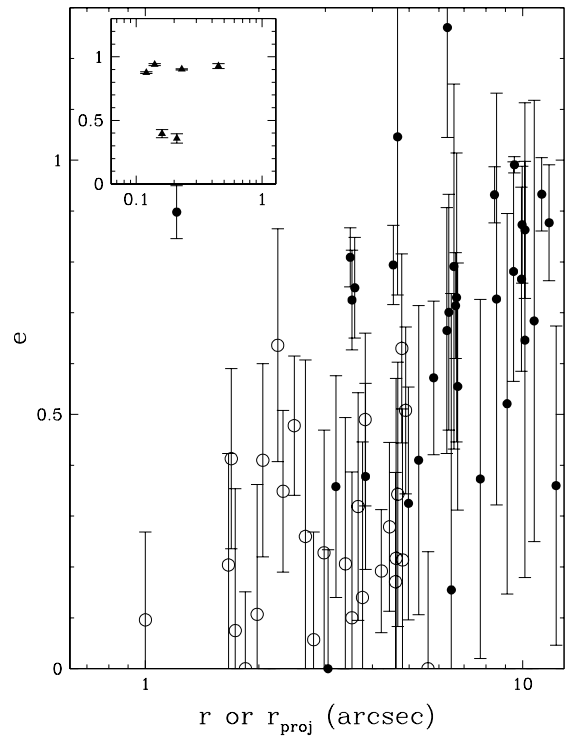}
\caption{The eccentricities of the young stars in the GC versus their
observed distance to the central MBH. The open circles represent
those stars having the measurements of the (three-dimensional)
distance ($r$). The solid circles represent those stars only having the
measurements of the projected distance ($r_{\rm proj}$). The inset
shows several S-stars which have the eccentricity estimates. The
figure shows a tendency that the outer stars have high eccentricities
(close to 1) while the inner ones (with exclusion of the S-stars) only
have moderate eccentricities. The data used here are adopted from
\citealt{Paumard06}.  Using the Spearman correlation analysis, we find
a quite strong correlation between the eccentricity and the projected
distance and the correlation coefficient between the eccentricity and
the projected distance is $R_S=0.58$ with $P_{R_S}=8\times 10^{-7}$.
} \label{fig:f3} \end{figure}

These features are consistent with the observed features of the orbits
partly listed in the items (a), (b) and (c) in \S~\ref{sec:intro}.  In
order to make a quantitative comparison with observations, we plot the
eccentricities of the young stars in the GC versus their observed
distance to the central MBH in Figure~\ref{fig:f3} (the data are
adopted from \citealt{Paumard06}).  This figure indeed indicates a
correlation that the stars in the outlying regions of GC have high
eccentricities (close to 1) while the inner ones only have moderate
eccentricities. This apparent correlation is reproduced in our model
as shown in panel (a) in Figures \ref{fig:distr} and
\ref{fig:distrinc90}.  Note that the low eccentricity of the test
particles in the inner region (with $r_f<0.15\pc$) preserved (panel a
in Figs.~\ref{fig:distr} and \ref{fig:distrinc90}), which appears
inconsistent with the observational range of $0.2-0.4$
(Fig.~\ref{fig:f3}). However, the eccentricities of the young stars at
the inner region may relax to the observed values due to the
interactions among the stars as shown in \citet{ABA07}, which are not
considered in \S~\ref{sec:model}. An N-body numerical simulation of
the model proposed in this paper would provide a quantitative
consistency check.
Our calculations in \S~\ref{subsec:migration} show that the inclination of the
stars in the outer region relative to the inner disk is less excited for a
nearly coplanar perturber orbit (only up to $\sim50\arcdeg$) than for an
overhead perturber orbit. The inclination angles of the outer stars are not
well determined in observations, but if they have higher inclinations (e.g.,
$\sim110\arcdeg$ in item a in the \S~\ref{sec:intro}), the perturber would be more
likely to be on an overhead orbit.  Precise determination of the inclination of
the stars in the outer region will be important to distinguish the inclinations
of the perturber.

In the in-spiraling-cluster scenario (briefly discussed in
\S~\ref{sec:intro} to account for the origin of the young stars), we
note that, within a limited dispersion, the eccentricities of the
tidally disrupted stars are usually comparable to that of the
in-spiraling star cluster.  The dominant component of the velocity
vector of the dispersed stars is their co-moving motion with the
cluster and the velocity of the stars relative to the center of the
cluster is relatively small \citep{BH06}. Therefore, it is unlikely
that the observed eccentricity distribution of the young stars shown
in Figure~\ref{fig:f3} can be produced by a disrupted single
in-spiraling young star cluster. In order to reproduce the observed
velocity distribution, the disruption of a second hypothetical 
in-spiraling young star cluster may be required (e.g., \citealt{BH06}).
Such a scenario seems contrived because the coeval nature of the
young stars also requires these clusters to be formed simultaneously.

The model we analyzed here is different from that two in-spiraling star
cluster scenario. In our model, all the young stars are initially in a
single disk and then the orbits of the young stars in the outer region
of the disk are perturbed by an inward-migrating (dark) star cluster
or an IMBH onto high eccentricity and high inclination orbits.

Figures \ref{fig:distr}, \ref{fig:distrinc90} and \ref{fig:distrinc30rp0.3}
(e) and (f) indicate one 
of the features of our
model that because they are captured into mean-motion resonance, some
test particles have nearly the same semimajor axes with widely
different eccentricities.  The particles captured into the resonance
are likely to have close encounters with each other, which should
modify the orbital distribution obtained here.  Our preliminary N-body
results show that some of the particles captured into the mean-motion
resonance may then be scattered out of the resonance.
Nevertheless, a substantial number of particles still attached to the
resonance.  The main conclusion derived in Figs.~\ref{fig:distr} and
~\ref{fig:distrinc90} are not expected to change significantly.  The
existence of resonant stars, if can be extracted from the
observational data, would not only provide an important check for our
model but also strongly signify the possible presence of an IMBH
in the GC.

Our simulation results show that the inclination angles of the young stars at
the outer region of the initial disk may be excited up to different values as
the perturber migrating inward.  The excited stars cannot form a coherent
secondary disk in this model.  The formation of a secondary disk is also
challenging in the in-spiraling cluster scenario \citep{BH06}. In any event,
observational evidence of the existence of a secondary disk remains
controversial.

\section{Conclusions} \label{sec:conclusion}

We studied the dynamical evolution of a young stellar disk surrounding
the MBH in the GC and perturbed by an inward migrating IMBH or a
(dark) star cluster hosting an IMBH. Our numerical simulations show that the
orbits of the young stars in the disk may be significantly modified
from the outside in by the perturber. If the perturber has a low
inclination angle to the initial disk, its migration from a region
outside to $\sim 0.15\pc$ may excite the eccentricities (up to unity)
and the inclination angles of stars in the outer (0.2--0.5\pc) regions
by capturing them into its mean-motion and secular resonances, forcing
them to migrate with it, and/or closely encountered with them.  Stars 
interior to this region preserve their
initial coplanar structure. The overall dynamical distribution of the stars
reproduces that observed on sub-parsec scale around the GC, i.e.,
an inner disk surrounded by a torus of highly eccentric and inclined
stars. If the perturber migrates to $\sim 0.3\pc$ on an orbit which is nearly
perpendicular to the initial disk, the stars in the outer regions can
still be excited to highly eccentric and inclined orbits through
resonant capture and nodal precession induced by and close encounters
with the perturber. The inclinations can be excited to be in a larger
range in the perpendicular case than those excited in the low-inclination
case. In the perpendicular case, the stars in the inner region can
also retain their low eccentricities and remain on the initial
disk. These results reproduce many features of the observed orbital
distribution of sub-parsec young stars. Note that the predicted position
for the perturber here is an instantaneous value, and the perturber does
not necessarily stop there but continues to migrate inward under the action
of dynamical friction with time going on.

Further measurements of the orbital parameters of the young stars in
the GC (e.g., semimajor axes, eccentricities, and inclination angles)
would provide important tests for the model proposed in this paper.
It would also be useful to distinguish various scenarios proposed to account
for the formation of the young stars.  An important confirmation for
the model proposed in this paper is the discovery of a perturber
[either an IMBH or a (dark) cluster hosting an IMBH] with a mass in the range $3\times
10^3-3\times 10^4\msun$ at $r\sim0.15-0.3\pc$ with an orbital plane which
is either nearly parallel or perpendicular to the inner disk plane.

\acknowledgments 
We have benefited from discussions with Scott Tremaine.
This work is supported by NASA (NAG5-12151, NNG06-GH45G),
JPL (1270927), NSF(AST-0507424). 
Q.Y. acknowledges initial support from NASA through Hubble Fellowship grant
HST-HF-01169.01-A awarded by the Space Telescope Science Institute, which is
operated by the Association of Universities for Research in Astronomy, Inc.,
for NASA, under contract NAS 5-26555.

\end{document}